\documentclass [11pt,a4paper]{article}
\usepackage{jcappub}

\usepackage{latexsym,epsfig,amssymb, amsmath,nicefrac}

\usepackage{color}

\usepackage[makeroom]{cancel}
  \usepackage{latexsym}
  \usepackage{epsf}
  \usepackage{amssymb}
  \usepackage{graphicx}
  \usepackage{amsmath}
  \usepackage{amsmath,amssymb,amsthm}
  \usepackage{verbatim}
  \usepackage{hyperref}

\def\tb{\tilde{\beta}}
\def\tNs{\tilde{N}_\star}
\def\tN{\tilde{N}}

\def\bi{\begin{itemize}}
\def\ei{\end{itemize}}
\def\be{\begin{equation}}
\def\ee{\end{equation}}
\newcommand{\bea}{\begin{eqnarray}}
\newcommand{\eea}{\end{eqnarray}}

\def\lsim{\mathrel{\mathop
  {\hbox{\lower0.5ex\hbox{$\sim$}\kern-0.8em\lower-0.7ex\hbox{$<$}}}}}
\def\gsim{\mathrel{\mathop
  {\hbox{\lower0.5ex\hbox{$\sim$}\kern-0.8em\lower-0.7ex\hbox{$>$}}}}}

\title{Can CMB data constrain the inflationary field range?}

\author[a]{Juan Garcia-Bellido,}
\author[b]{Diederik Roest,}
\author[b]{Marco Scalisi}
\author[b]{and Ivonne Zavala}

\affiliation[a]{Instituto de F\'isica Te\'orica IFT-UAM-CSIC, Universidad Aut\'onoma de Madrid,
C/ Nicol\'as Cabrera 13-15, Cantoblanco, 28049 Madrid, Spain}
\affiliation[b]{Centre for Theoretical Physics, University of Groningen, \\ Nijenborgh 4, 9747 AG Groningen, The Netherlands}

\emailAdd{juan.garciabellido@uam.es}
\emailAdd{d.roest@rug.nl}
\emailAdd{m.scalisi@rug.nl}
\emailAdd{e.i.zavala@rug.nl}
\abstract{We study to what extent the spectral index $n_s$ and the tensor-to-scalar ratio $r$ determine the field excursion $\Delta\phi$ during inflation. We analyse the possible degeneracy of $\Delta \phi$ by comparing three broad classes of inflationary models, with different dependence on the number of e-foldings $N$, to benchmark models of chaotic inflation with monomial potentials.  
The classes discussed cover a large set of inflationary single field models.  We find that the field range is not uniquely determined for any value of $(n_s, r)$; one can have the same predictions as chaotic inflation and a very different $\Delta \phi$. Intriguingly, we find that the field range cannot exceed an upper bound that appears in different classes of models. Finally, $\Delta \phi$ can even become sub-Planckian, but this requires to go beyond the single-field slow-roll paradigm.}
\arxivnumber{1405.7399}

\begin{document}

\maketitle

\section{Introduction}

Cosmological inflation \cite{Guth,Linde81,AS82}, originally proposed as a natural explanation for the homogeneity and flatness of our Universe, has been put on very firm grounds thanks to the recent observations. Data indeed seem to confirm more strongly the inflationary paradigm as the leading mechanism to account for the origin of  the anisotropies in the Cosmic Microwave Background (CMB) radiation and, thus, the formation of the large scale structures. In particular, they  support two robust predictions of inflation, that is, a nearly scale invariant spectrum of density perturbations and a stochastic background of gravitational waves.

The Planck satellite \cite{Planck} has reported tighter constraints on the inflationary parameters, specifically for the spectral index $n_s$ whose value turns out to be
\be\label{infladata}
n_s=0.9603\pm0.0073. 
\ee
The BICEP2 collaboration \cite{bicep}, on the other hand, recently claimed the first measurement of the tensor to scalar ratio $r$ to be 
\be\label{t2s}
 r = 0.20^{+0.07}_{-0.05} \,,
\ee
with $r=0$ disfavored at $5\sigma$. This value is in slight tension with the reported Planck upper bound being $r<0.11$ at 95\% c.l., although foreground subtraction generically tends to decrease the reported value.\footnote{In 
ref.~\cite{Audren:2014} the authors point out a mismatch both between the pivot points of Planck and BICEP2 and their assumptions on $n_t$ (see also \cite{Ashoorioon:2014nta}). Taking these differences into account softens the tension between them by lowering the measurement \eqref{t2s} to $r=0.16^{\,+0.06}_{\,-0.05}$ and the discrepancy becomes of order 1.3$\sigma$ only.}

If confirmed as a primordial signal, such detection would definitively lead to impressive consequences. First of all, it would finally provide a strong evidence for the quantization of the gravitational interaction \cite{Ashoorioon:2012kh, KraussW}. Secondly, it would put dramatic constraints on the inflationary variables, ruling out a considerable number of cosmological models. For example, according to the Lyth bound \cite{Lyth}, one of the strongest implication of such large value of $r$ would be a super-Planckian excursion for the inflaton field.

Both data sets come from CMB observations which probe a region corresponding to horizon crossing at present around 50 to 60 e-foldings before the end of inflation. This is quantified by the number of e-folds $N$, defined as
\be\label{eqN}
N= N_\star - N_e \equiv\log \frac{a_e}{a_\star} = \int_{t_\star}^{t_e} H \,dt \,,
\ee
where the subscripts $\star$ and $e$ refer, respectively, to the scales that cross the Hubble radius during inflation
and are now crossing back into our present horizon, and those at the end of inflation, while $a$ is the scale factor and $H$ is the Hubble parameter, which is integrated over the whole inflationary period. A lower limit on $N \gsim 50$ can be set by the temperature of reheating, given the BICEP2 results \citep{Dai:2014jja}. On the other hand, there is no compelling reason to assume the number $N$ has an upper bound on the amount of exponential expansion of the Universe; in fact, it seems natural that inflation extends a long way further into the past than the portion that we can observe (see \cite{RemCar} for a recent study on this topic).

The above argument seems to suggest $1/N$ as a natural small parameter to expand our cosmological variables. Moreover, the percent-level deviation from unity of $n_s$ together with the present measurements of $r$ support even more such idea since their values can be naturally accommodated within a perturbative $1/N$ expansion. Considering large values of $N$ proves to be a powerful tool in order to organize different inflationary models just in terms of their cosmological predictions. As shown in previous works \cite{Mukhanov,Roest,GarRoe}, it is possible to identify a number of classes where physically different inflationary scenarios would predict the same values of $n_s$ and $r$ in the leading approximation in $1/N$ (see also \citep{Boyanovsky:2005pw}). In some classes it can be shown that subleading corrections are irrelevant from the observational point of view.

The inflationary phase can be specified fully in terms of $N$ \cite{GarRoe} and this has the direct advantage to go beyond an explicit description of the microscopic mechanism generating the accelerated expansion and the deviation from a scale invariant spectrum.

In the light of the recent data release, it seems worthwhile to examine features of a variable crucial for the construction of inflationary models at high energies, namely the inflaton field range $\Delta\phi$. Assuming that quadratic inflation provides a good fit to both the Planck and BICEP2 data, does this imply that the field range is necessarily identical to that of a massive free field? In the present paper we address this question and extend the analysis also to other chaotic monomial inflation scenarios, which naturally provide a large value of $r$. However, our approach can be applied to any other model of inflation with given predictions for\footnote{In this paper we do not consider the running of the spectral tilt; more precise measurements of the running can be used to narrow the admissible window for the inflationary range.} $(n_s, r)$. In the chaotic example, we prove the existence of an upper bound on $\Delta\phi$ and the total number of e-folds $N$. Finally, we show that in some cases it is possible in principle to have sub-Planckian field ranges, if one goes beyond the requirements of single-field and/or slow-roll inflation \cite{Antusch}.

The paper is organized as follows. In section \ref{sec:chaotic}, we discuss the large-$N$ behavior of models with natural large $r$, namely chaotic inflation scenarios with monomial potentials~\cite{Linde:1983gd}. These will be regarded as benchmark models throughout the paper. Then, in section \ref{sec:range}, we analyze three different classes of models with specific dependence on $N$, which describe several inflationary models in the literature. We point out connections among them and  present the main results on the degeneracy of the field range together with its upper bound.  We summarise our results and present an outlook in the discussion \ref{sec:disc}. Throughout the paper we set the reduced Planck mass $M_P=1$.

\section{Chaotic inflation as benchmark}
\label{sec:chaotic}

The classification of inflationary models at large-$N$ \cite{Mukhanov,Roest, GarRoe} can be elegantly performed in terms of the Hubble parameter $H$, which encodes the accelerated expansion and its evolution as a function of the number of e-folds (note that $N$ decreases as time progresses in our conventions). This is nicely described  in terms of the Hubble flow functions $\epsilon_n$ defined iteratively as \cite{STEG,STE}
 \be\label{flow}
  \epsilon_{n+1} = \frac{d \log|\epsilon_n| }{dN}\,. 
 \ee
 The first of these quantities is identical to the Hubble parameter, $\epsilon_0 = H$. The inflationary observables, in particular the spectral index of the scalar density perturbations and the tensor-to-scalar ratio, can then be compactly expressed as
\be\label{inflaparam}
n_s =1+\epsilon_2 -2\,\epsilon_1,    \qquad  r=16\,\epsilon_1 \,.
\ee
In order to connect these to CMB observations, one needs to evaluate these quantities at horizon crossing, denoted by $N_\star$.

The formulation in terms of $N$ does not depend on the mechanism that drives the inflationary period, nor on the details of the underlying microscopic model. However, strictly within the slow-roll approximation, the Hubble flow functions are equivalent to the flow parameters in terms of the potential $V$ for a scalar field $\phi$, namely:
   \be\label{flowSR}
  \epsilon_0 = V^{1/2} \,, \quad \epsilon_1 = \epsilon \,, \quad \epsilon_2 = - 4 \epsilon + 2 \eta \,,
 \ee
where  the slow-roll parameters are defined as
\begin{align}\label{SRP}
  \epsilon = \frac12 \left( \frac{V_\phi}{V} \right)^2 \,, \quad \eta = \frac{V_{\phi \phi}}{V} \,.
 \end{align}
The link between these two equivalent formulations is provided by the usual relation
\be\label{dphi1}
\frac{d \phi}{dN} = \sqrt{2 \epsilon_1} \,.
\ee 
The  Ansatz $\epsilon (N)$ therefore fully determines the scalar potential of a corresponding inflationary model, either in terms of a scalar field $\phi$ with canonical kinetic terms, or by the Lagrangian
 \begin{align}
 \mathcal{L}  = \sqrt{-g}\,[ \tfrac12 R -  \epsilon(N) (\partial N)^2 - V(N) ] \,,
 \end{align}
when interpreting the number of e-foldings as a field $N$. Generically, the functional form of the potential $V$ will be very different when expressed in terms of $\phi$ or in terms of $N$.

Motivated by the recent cosmological data, we consider chaotic inflation scenarios as benchmark models for the following study. However, note that other models can be straightforwardly studied following the same reasoning. Chaotic scenarios are usually characterized by monomial potentials when expressed in terms of the canonical scalar field $\phi$. Further, they naturally lead to a large value of $r$ together with a super-Planckian excursion of the inflaton field. In a large-$N$ description, the first three Hubble flow functions turn out to be
\be\label{chaopar}
\epsilon_0= h N^\beta\,,    \qquad \epsilon_1= \frac{\beta}{N}\,,    \qquad  \epsilon_2=-\frac{1}{N}\,,
\ee
where $h$ is an integration constant and $\beta$ is related to the specific universality class.

The description in terms of $N$ is exact for these models (there are no subleading corrections) and hence captures all of their fundamental features. However, even if there would be subleading corrections, e.g.~at the level $1/N^2$, observables calculated at horizon exit, such as $n_s$ and $r$, will be observationally insensitive to these (as they are too much suppressed for $N \gsim 50$). Therefore, these are universal predictions of entire classes of models that agree in the large-$N$ limit. 

The same universality holds for the inflaton range. In the case of chaotic models with parameters \eqref{chaopar}, the inflaton excursion $\Delta \phi$ will be basically determined just by the leading term in $N$ \cite{GBRSZuni} through Eq.~\eqref{dphi1}.

As these models receive most of their e-foldings at large-$N$, one can safely assume that restricting to the leading term of $\epsilon_1$ is a very good approximation over the relevant part of the inflationary trajectory. The expression for the inflaton field range will therefore read
\be\label{DPchaotic}
 {\Delta \phi}_c = 2 \sqrt{2\beta} \left( N_\star ^{1/2}-N_e^{1/2} \right)\,,
 \ee
where the subscript $c$ is added in order to refer more easily to the benchmark field excursion of monomial models throughout the paper. Further, $N_e=\beta$, when assuming that inflation ends at $\epsilon_1=1$, and $N_\star$ is found through \eqref{eqN}. 

With the above relations, potentials of the type $V (\phi)= \lambda_n \phi^n$ will keep monomial form even when formulated in terms of $N$, namely $V(N)=h^2 N^{2\beta}$, and vice versa. The relation between the two power coefficients reads
\be
\beta =\frac{n}{4}\,,
\ee
and can be found by using Eq.~\eqref{dphi1}. As an explicit example, a quadratic potential  corresponds to $\beta=1/2$ and an inflationary period of $N=60$ leads to $\Delta \phi  \simeq 14.14$. Of course, this is identical to the value of $\Delta \phi$ calculated through the scalar potential $V$, within the slow-roll paradigm.

\section{Degeneracy of the field range}
\label{sec:range}

We now discuss the field range in different classes of models. In particular, we are interested in exploring the correspondence between a specific point in the $(n_s,r)$ plane and the values of $\Delta \phi$. We will prove that it is possible to have exactly the same cosmological predictions, in terms of the scalar tilt and the amount of gravitational waves, while the field excursion may vary over several orders of magnitude. 

We analyze three classes of inflationary models with a specific dependence on $N$ for the Hubble flow parameters. Such classes, discussed at length in \citep{GarRoe}, reproduce the large-$N$ behaviour of most of the inflationary models available in the literature.

As a first case, we  discuss the so-called {\it perturbative} class, characterized by a leading term in $\epsilon_1$ scaling as $1/N^p$, with $p$ being a constant positive coefficient. Then, we analyze models where {\it logarithmic} terms, such as $\ln^q(N+1)$,  appear in the leading part of $\epsilon_1$. In a  third class of models,  we consider the parameter $\epsilon_1$ having a {\it non-perturbative} form, in the limit at large-$N$, of the type $\epsilon_1 \sim \exp(-cN)$. We will consider the possibility of letting the total number of e-folds $N$ vary over a certain interval which is related to reheating details of the specific model. Interestingly, we find an {\it upper bound} on $\Delta\phi$ and the total number of e-folds which sets connections among the three classes of models considered.

As final part of our analysis, we focus on the logarithmic class and we explore the possibility of playing  with the power coefficient $q$, while keeping $N$ fixed. This is an alternative way to get the same predictions of quadratic inflation, while having quite different values for the inflaton range. We will consider the possibility of going beyond single-field and/or slow-roll inflation and getting a sub-Planckian $\Delta \phi$.

Throughout the paper, we assume that the inflationary parameters $\epsilon_1$ and $\epsilon_2$ of each class are exact over the whole inflationary trajectory, as it happens for chaotic scenarios. In several cases, this may be a very good approximation and may capture most of the essential properties of the models falling into the specific universality classes. Anyhow, we will take advantage of a formulation purely in terms of $N$ and extract the information we are interested in, without referring to the particular form of the scalar potential $V(\phi)$. In fact, for any specific parametrization of each class, the latter may be very complicated when expressed in terms of the canonical scalar field $\phi$.

In what follows, the benchmark will be the value of $\Delta \phi$ for chaotic models corresponding to a quasi exponential expansion of $N=60$. This sets
\be \label{Nstpert}
N_\star=\beta + 60\,,
\ee
 as corresponding to horizon exit.
Moreover, all symbols with a tilde  will be reserved for the classes being examined, while the benchmark models will have no tilde.

\subsection{Perturbative class} \label{PClass}

We start considering the possible degeneracies within the perturbative class of models. In this case, the relevant Hubble flow parameters for determining the observational data  have the following $N$-dependence:
\be\label{pertpar}
\epsilon_1= \frac{\tb}{N^p}\,,    \qquad  \epsilon_2=-\frac{p}{N}\,.
\ee
The case discussed in sec. \ref{sec:chaotic} is easily recovered for $p=1$ and $\tb=\beta$.

\begin{figure}[htb]
\hspace{-3mm}
\begin{center}
\includegraphics[width=8.5cm]{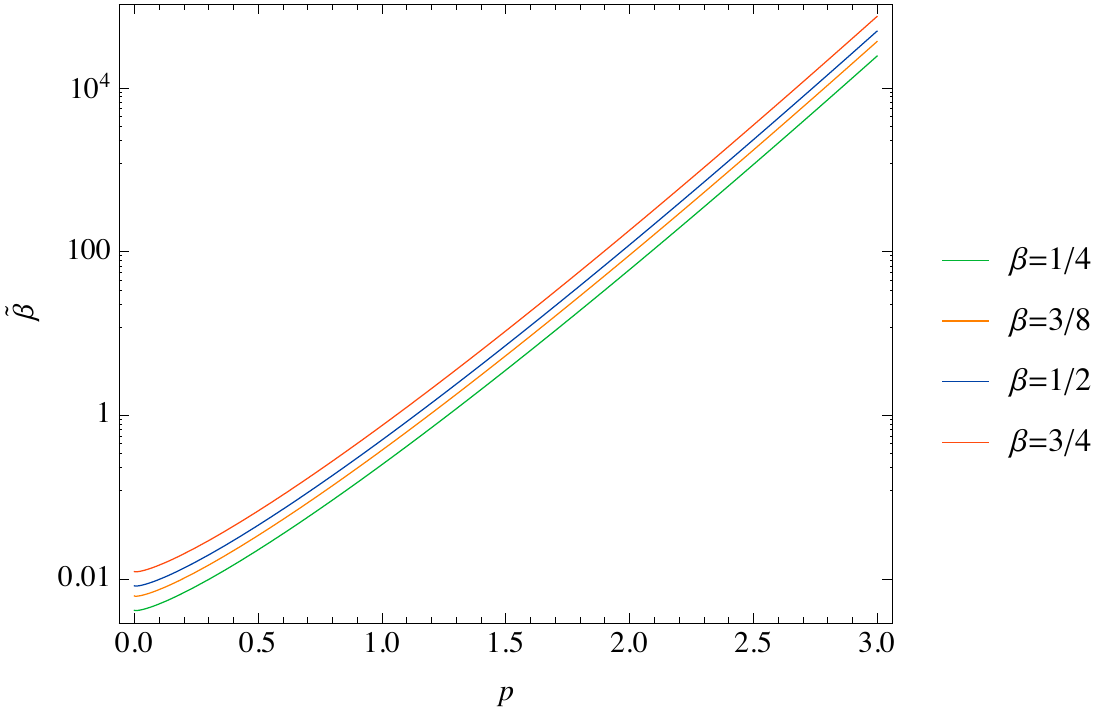}
\caption{Behavior of $\tb$ as function of $p$ in a log-plot. It generally blows up for $p\gtrsim2$, where the  perturbative model should be highly fine-tuned in order to reproduce the same $(n_s,r)$ of chaotic scenarios. The four lines correspond to the same observational predictions of models with potential of the type $V = \lambda_n \phi^n$, with $n$ respectively equal to $1$, $3/2$, $2$ and $3$. }\label{FIGbeta}
\end{center}
\end{figure}

We would like to reproduce the same $n_s$ and $r$ of the benchmark chaotic model through a generic pertubative model with $p\neq1$. This translates into equating both $\epsilon_1$ and $\epsilon_2$ of \eqref{chaopar} to the functions \eqref{pertpar} at horizon exit, respectively at $N_\star$ and $\tNs$. As result, we have the following relations:
\bea\label{btper}
\tb &=& \beta\ \frac{\tNs ^p }{N_\star} \,, \qquad
p =  \frac{\tNs}{N_\star}\,.
\eea 
This allows  to express $\tb$ as
\be\label{beta}
\tb = \beta\ p^p N_\star^{p-1} \,,
\ee
where $N_\star$ is given by \eqref{Nstpert}. Eq. \eqref{beta} gives us an estimate of how fine-tuned the model is in order to reproduce the same predictions of the chaotic models. Curiously, for any  $\beta=\mathcal{O}(1)$ (corresponding to different chaotic models), the corresponding perturbative model will start to be severely fine-tuned in the region $p>2$, as is shown in Fig.~\ref{FIGbeta}.

\vspace*{-0.5mm}

\begin{figure}[htb]
\hspace{-3mm}
\begin{center}
\includegraphics[width=8.5cm]{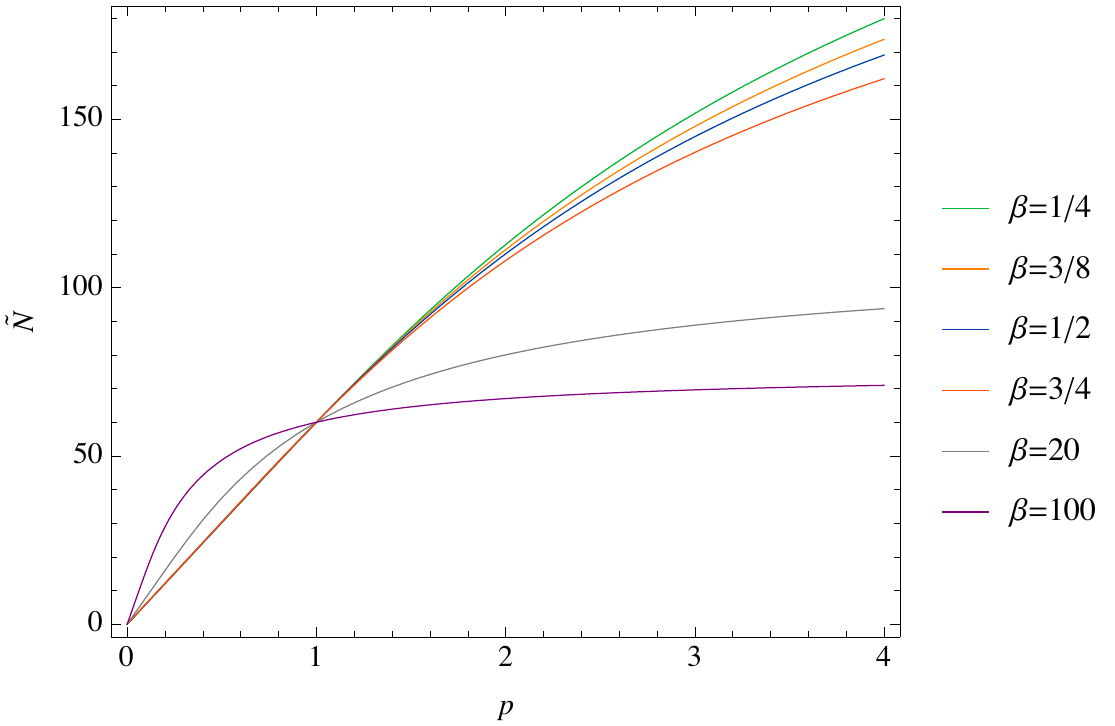}
\caption{The total number of e-folds $\tN$ as function of $p$ for models belonging to the perturbative class. Lines follow a linear relation for low values of $p$ and $\beta=\mathcal{O}(1)$. For larger values of $\beta$, lines have different behaviors, as shown by the grey and purple lines. In this case, the unique intersection with the value $p=1$ becomes evident.} \label{FIGnpert}
\end{center}
\end{figure}

Demanding that inflation ends at $\epsilon_1=1$ turns into 
\be\label{Neper}
\tilde{N}_e = \tNs - \tilde{N} = \tb ^{1/p} \,,
\ee
where the total number of e-foldings $\tilde{N}$ in principle could span a range of different values related to reheating properties of the model. Using \eqref{btper}, Eq.~\eqref{Neper} gives us the functional form of the total number of e-folds $\tilde{N}$ as a function of $p$, for any $\beta$, that is
\be\label{Nper}
\tilde{N} = p \,N_\star \left[1-  \left(\frac{\beta}{N_\star}\right)^{1/p}\right]\,.
\ee
A period of inflation $\tilde{N}=60$ necessarily corresponds to $p=1$, which is the benchmark of our analysis. For any other value of $\tilde{N}$, there exist several possibilities with $p\neq1$, reproducing exactly the same predictions of chaotic models, while having a viable mechanism to end inflation ($\epsilon_1=1$). Nevertheless, in the region $p<2$ and for $\beta=\mathcal{O}(1)$, solutions in $p$ are highly close together and they follow a linear relation, as shown in Fig.~\ref{FIGnpert}.

\begin{figure}[htb]
\hspace{-3mm}
\begin{center}
\includegraphics[width=7.6cm]{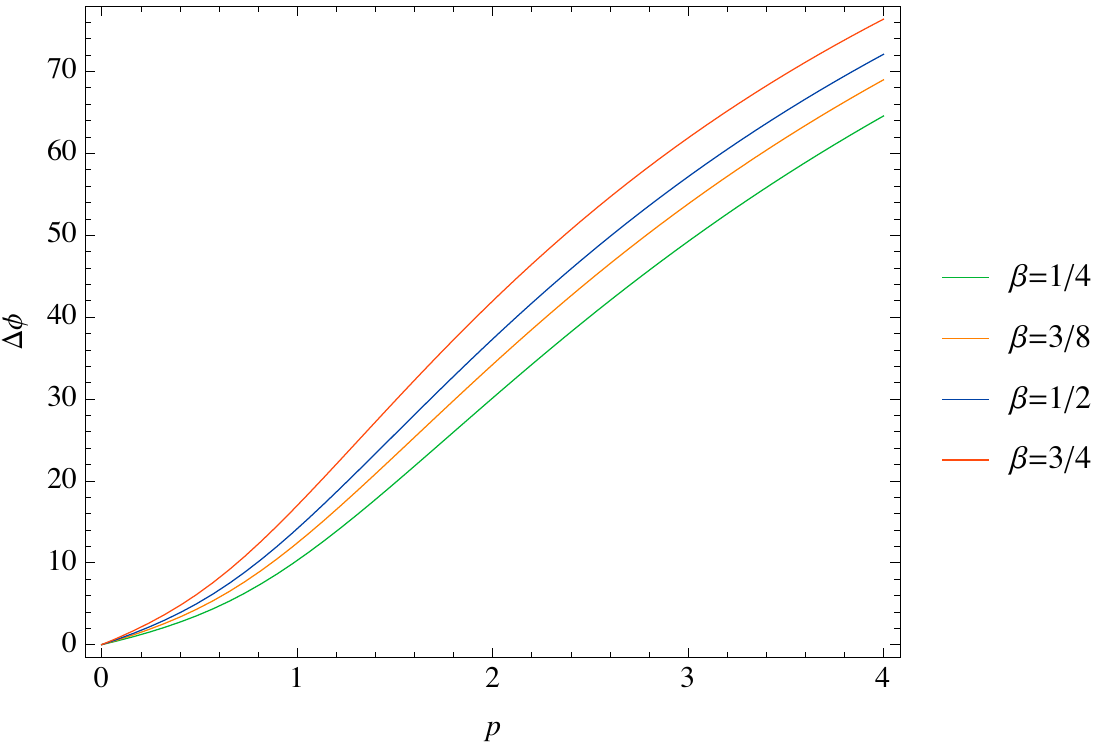}
\includegraphics[width=7.6cm]{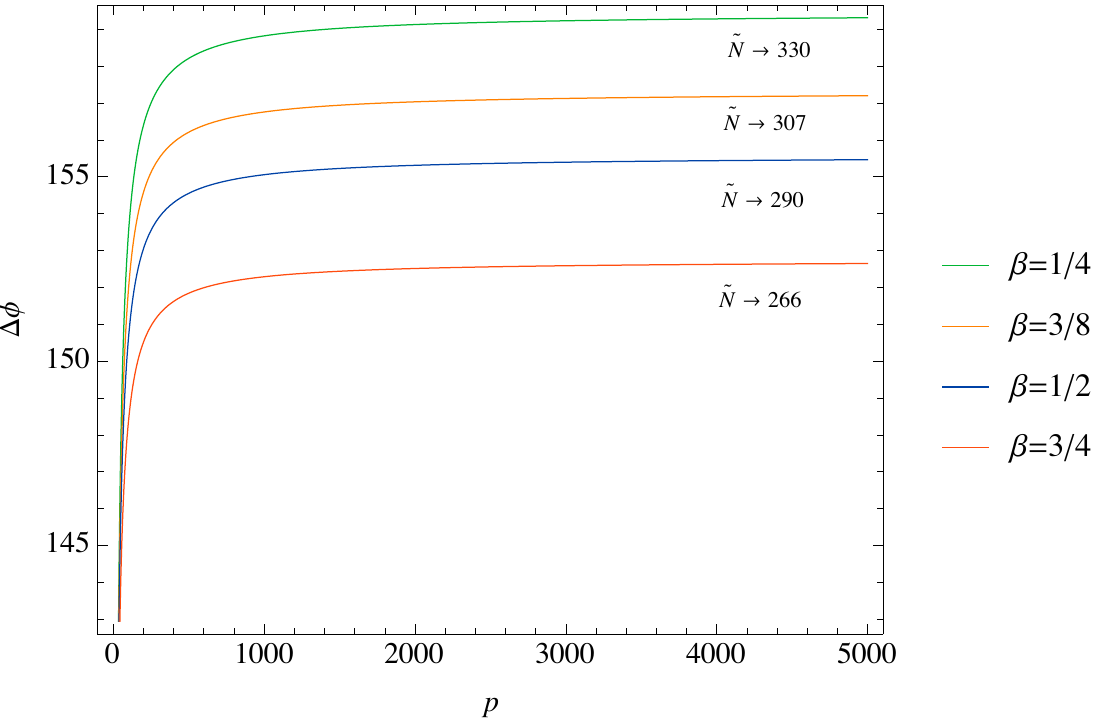}
\caption{The inflaton range $\Delta\phi$ as function of $p$ in two different limits for the pertubative class of models. For low values of $p$ (no fine-tuning), $\Delta\phi$ may vary over a range related to the total number of e-folds $\tN$. In the large $p$ region, the upper bounds on $\Delta\phi$  becomes evident.} \label{FIGdeltapert}
\end{center}
\end{figure} 

The inflaton range can be easily computed by integrating Eq.~\eqref{dphi1} and one obtains
\be
\Delta \phi = \frac{2\sqrt{2\tilde\beta}}{2-p}  \left(\tilde N_\star^{1-\frac{p}{2}} -\tilde N_e^{1-\frac{p}{2}}\right)\,.
\ee
This latter can be written as function of $p$, for any value of $\beta$. By substituting \eqref{btper}, one gets
\be\label{DPpert}
\Delta \phi = 2\sqrt{2\beta} \frac{p}{2-p}  \left[ N_\star^{\frac{1}{2}}- \beta^{\frac{2-p}{2p}}N_\star^{\frac{p-1}{p}}\right]\,.
\ee
Figure \ref{FIGdeltapert} shows the main results on the inflaton range, given by \eqref{DPpert}, for models belonging to the perturbative class. At this point, we identify the two regions and get the following conclusions:
 \begin{itemize}
 \item For small values of $p$, the inflaton excursion $\Delta\phi$ is a continuously increasing function. It has a typical dependence $p/(2-p)$ in the region $p\lesssim1$, where the first term of \eqref{DPpert} dominates over the second one; it has a mild transition for $1\lesssim p \lesssim2$, while it starts to show a really different behavior in the region\footnote{The value $p=2$ is special as the two contributions of Eq.~\eqref{DPpert} become the same while the factor $p/(2-p)$ blows up.} $p>2$. The field range covers a wide spectrum of values depending on the total number of e-folds $\tN$ of this perturbative class. As a consequence, it can be quite different from the corresponding chaotic one, which is given by $p=1$. In particular, we can reproduce the same values $(n_s,r)$ of a quadratic potential with $N = 60$ and still have a $\Delta\phi$ running from  5 to 32, in Planck units, corresponding to $\tN$ approximately between 30 and 100. Note that $p$, as well as $\Delta\phi$, cannot be arbitrarily small as we need a minimum amount of exponential expansion, quantified by $\tN$.

 \item For large values of $p$, the inflaton range approaches a constant value, setting an {\it upper bound} on $\Delta\phi$ for each specific value of $\beta$.  This can be  seen explicitly by taking the limit of  \eqref{DPpert} for $p\to\infty$, this becomes:
  \be\label{dfbound}
 \Delta \phi \to 2\sqrt{2}N_\star^{1/2}\left[N_\star^{1/2} -\sqrt{\beta}\right]\,.
\ee
 This corresponds to an upper bound also on $\tN$, as can be seen again by taking the limit for $p\to \infty$ of equation \eqref{Nper}, which gives:
 \be\label{Nbound}
 \tilde N \to  N_\star \ln \frac{N_\star}{\beta} \,.
 \ee
 This limit  cannot be appreciated in Fig.~\ref{FIGnpert}, given the reported limited range of $p$. Plugging the values of the parameters for quadratic inflation into \eqref{dfbound}  and \eqref{Nbound}, one gets the approximate bounds\footnote{Such large values of $\tilde N$ are not necessarily realistic (see e.g.~the discussion in \cite{Liddle:2003as} for an upper estimate); nevertheless, it is interesting to study the behaviour of the field range for such models.}
 \be
\Delta\phi\to 155.56\,, \qquad \tN\to 290\,.\label{boundquadr}
\ee
Curiously, the hierarchy of ranges is inverted with respect to the one present at small $p$, as it is clear by comparing the two pictures of Fig.~\ref{FIGdeltapert}: at higher values of the tensor-to-scalar ratio $r$, we have smaller ranges. 
 
 We will see that the bounds for $\Delta\phi$ and $\tN$ found here are recovered in the next two cases we consider in  sec.~\ref{Log} and \ref{NP}, within the analysis of the logarithmic and non-perturbative classes of models. 
\end{itemize}

\subsection{Logarithmic class}\label{Log}

As a second case, we consider models with a first subleading correction to the Hubble flow parameters. While still neglecting higher order $1/N$ terms, one can imagine including a logarithmic dependence on $N$ such as
\be \label{logep1}
\begin{aligned}
\epsilon_1 &= \frac{\tb}{N^p\ln^q(N+1)}\,,\\
\epsilon_2 &= -\frac{p}{N} -\frac{q}{(N+1)\ln(N+1)} \,.
\end{aligned}
\ee
Inflationary models having similar dependence can be found e.g. in \cite{Roest}.

As in the previous case, in order to mimic the observational predictions of chaotic models in terms of ($n_s,r$), we equate ($\epsilon_1$, $\epsilon_2$) of \eqref{chaopar} to \eqref{logep1} at horizon exit, respectively at $N_\star$ and $\tNs$. As result, we obtain
\bea\label{deg2}
&&\tb = \beta\ \frac{\tNs^p\ \ln^q(\tNs +1)}{N_\star} \\
&&  q= \left(\frac{1}{N_\star}-\frac{p}{\tilde N_\star}\right)(\tilde N_\star + 1) \ln(\tilde N_\star +1)\label{deg3}
\eea
where  $\tNs = \tilde N_e + \tN$ and $\tilde N_e$ is determined by the condition $\epsilon_1=1$:
\be\label{Nf2}
\frac{\tilde \beta}{\tilde N_e^p \ln^q(\tilde N_e +1)} =1\,.
\ee

We follow the same approach as in the perturbative case and allow $\tilde N$  to vary as function of $p$, while fixing $q$. The range of the inflaton $\Delta \phi$ can be determined by integrating \eqref{dphi1} as before. However, we have to rely on numerics as obtaining an analytic expression both for $\tN$ and $\Delta\phi$ turns out to be not as trivial as in the previous case. For this reason, we restrict our analysis just to the benchmark of a quadratic potential, namely just to $\beta=1/2$.

\begin{figure}[htb]
\hspace{-3mm}
\begin{center}
\includegraphics[width=8.5cm]{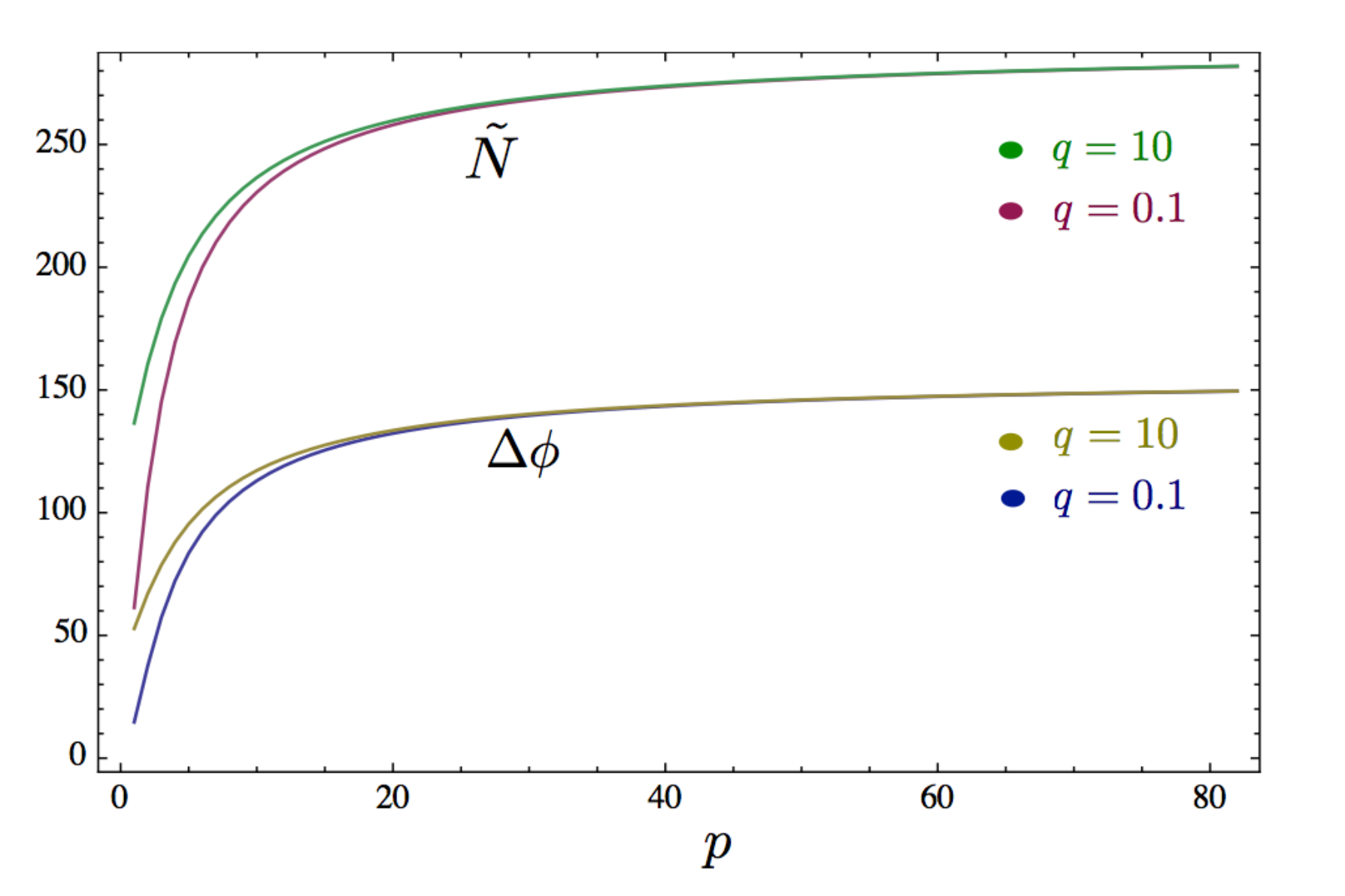}
\caption{Field range $\Delta\phi$ and total number of e-folds $\tilde N$ as functions of $p$ in the logarithmic class, for two fixed values of $q$. The lines correspond to the same predictions of quadratic inflation ($\beta=1/2$). The bounds on $\Delta \phi$ and $\tN$ can be appreciated at large values of $p$.}
\label{FRLogdNJoined}
\end{center}
\end{figure}

The results for the field range and the total number of e-folds are summarized in Fig.~\ref{FRLogdNJoined}, for two different values of $q$. As we can see, for large values of $p$, we recover exactly the same bounds \eqref{boundquadr} found within the analysis of the perturbative class. This is a remarkable result, though it may be understood from the large $p$ behaviour of $\epsilon_1$. In this limit, $\tN$ also increases and hence subleading terms, in $\epsilon_n$ for $n \geq 2$, will be increasingly irrelevant. The two lines in Fig.~\ref{FRLogdNJoined}, corresponding to different values of $q$, do differ for smaller values of $p$. However, they show identical behavior when $p$ increases, which correspond to a large-$\tN$ limit.

\subsection{Non-perturbative class}\label{NP}

As a third class, we consider models with Hubble flow functions such as
\be\label{npertpar}
\epsilon_1= \ e^{-2cN}\,,    \qquad  \epsilon_2=-2 c\,,
\ee
where $c$ is a constant. Note that we are not including any coefficient for $\epsilon_1$ as this can be set equal to one by a shift in $N$.

We proceed as in the previous cases by equating ($\epsilon_1$, $\epsilon_2$) of \eqref{chaopar} to the functions \eqref{npertpar} at horizon exit, in order to reproduce the same observational predictions of chaotic inflation models. We get the following relations:
\bea\label{btnper}
\tNs &=& \frac{1}{2c} \ln\frac{N_\star}{\beta} \,, \qquad
c =  \frac{1}{2N_\star}\,. 
\eea 
Moreover, imposing that inflation ends at $\epsilon_1=1$ translates into
\be\label{Nenper}
\tilde{N}_e = \tNs - \tilde{N} = 0 \,,
\ee
which can be manipulated, using \eqref{btnper}, in order to get the following condition on the total number of e-foldings:
\be \label{TNnper}
\tN = N_\star \ln\frac{N_\star}{\beta}\,,
\ee
expressed just in terms of parameters of the benchmark models, where $N_\star$ is given by \eqref{Nstpert}. Eq.~\eqref{TNnper} fixes uniquely the total amount of exponential expansion required to give the same $(n_s,r)$ of the chaotic scenarios, with parameter $\beta$, and to end inflation via the condition $\epsilon_1=1$. Note that this coincides exactly with the large-$p$ limit of the perturbative case, namely Eq.~\eqref{Nbound}. 

\begin{figure}[htb]
\begin{center}
\includegraphics[width=7.6cm]{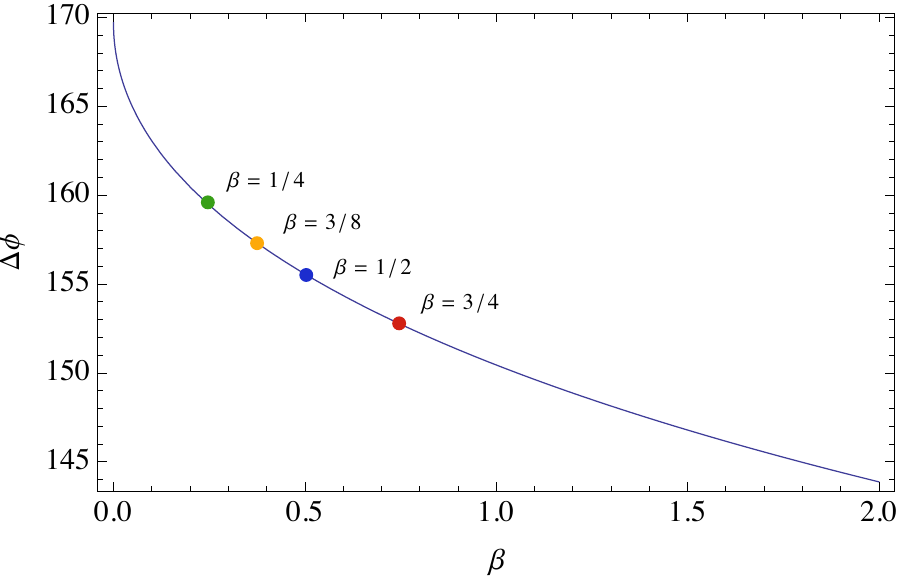}
\includegraphics[width=7.6cm]{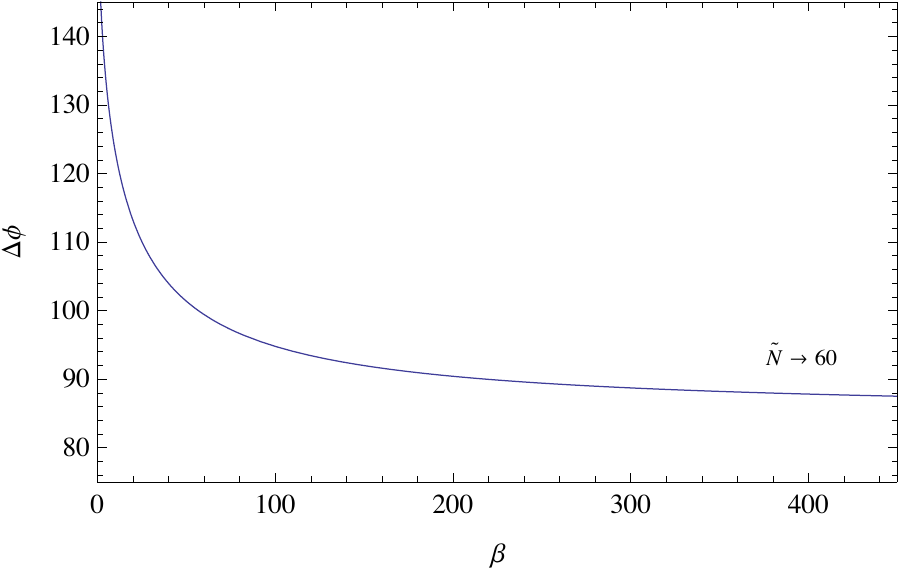}
\caption{The inflaton range $\Delta\phi$ as function of $\beta$ in two different limits for the non-perturbative class of models. For low values of $\beta$ (physical values for the tensor-to-scalar ratio $r$), $\Delta\phi$ is a decreasing function. The four coloured points correspond to the upper-bounds already found in sec.~\ref{PClass}  and \ref{Log}. In the large-$\beta$ limit, the inflaton range cannot arbitrarily decrease and approaches a lower-limit.} \label{FIGDeltaPhiNP}
\end{center}
\end{figure}

The inflaton range is given by integrating Eq.~\eqref{dphi1} between $\tN_e$ and $\tNs$:
\be\label{DPnper1}
\Delta \phi = \frac{\sqrt{2}}{c}  \left( 1-e^{-c\tNs}\right)\,.
\ee
The latter can be written just in terms of the benchmark parameters by using \eqref{btnper} and it reads
\be\label{DPnper}
\Delta \phi = 2\sqrt{2} \left(N_\star-\sqrt{\beta N_\star}\right)\,,
\ee
which yields the field range in terms of $\beta$. Note that this again coincides exactly with the large-$p$ limit of the field range in the perturbative case, that is \eqref{dfbound}. Fig.~\ref{FIGDeltaPhiNP} shows such functional dependence and the negative slope of the curve makes explicit the inversion of hierarchy of field ranges with respect to the one which naively one would expect. In fact, lower values of $r$ (lower values of $\beta$) will correspond to larger $\Delta\phi$. This is exactly the same finding for the upper bounds in the perturbative class of models. Such behavior becomes explicit once we express Eq.~\eqref{DPnper} in terms of the typical inflaton range ${\Delta\phi}_c$ for the chaotic models, given by Eq.~\eqref{DPchaotic}. The relation turns out to be:
\be
\Delta \phi = 2\sqrt{2} N - {\Delta\phi}_c \,,
\ee
where $N$ is the total number of e-folds for the benchmark chaotic models and, throughout our study, it is fixed to be equal to $60$. However, it is not possible to arbitrarily decrease $\Delta\phi$ even going to really large values of $\beta$. In fact, by taking the limit for\footnote{Such a limit is anyway not physical as it would correspond to an infinitely large amount of primordial gravitational waves.} $\beta\rightarrow\infty$ of \eqref{DPnper}, we obtain
\be
\Delta\phi \rightarrow \sqrt{2} N\,,
\ee
as can be seen in the second plot of Fig.~\ref{FIGDeltaPhiNP}, where $N=60$. This corresponds to a lower-bound on $\tN$ which, in the same limit, approaches the benchmark number of e-folds $N$, as it clear by taking the limit of \eqref{TNnper}. The field range of $\sqrt{2} N$ can then be understood from an $\epsilon_1$ parameter that is approximately equal to one during almost the entire inflationary period.

Within the non-perturbative models, it is then possible to mimic chaotic scenarios in terms of their cosmological observables $n_s$ and $r$. Nevertheless, both the total number of e-foldings $\tN$ and the field excursion $\Delta\phi$ are uniquely determined once we choose the power coefficient of the chaotic scenario, namely once we fix $\beta$. Curiously, the resulting values perfectly correspond to the upper-limits we found in the previous sections. In the specific example of quadratic inflation, that is for $\beta=1/2$, one obtains again $\Delta\phi\approx 155.56$ and $\tN\approx290$, as  expected from the discussion in sec.~\ref{PClass}  and \ref{Log}. 

Note, however, that this limit appears only in the large-$N$ limit of the non-perturbative class. We have discussed specific models of this class in \cite{GarRoe}. An example is natural inflation \cite{Freese:1990rb}, which has specific subleading corrections in addition to \eqref{npertpar}. In the limit of a large periodicity, corresponding to small $c$, this model asymptotes to quadratic inflation and therefore has the same field range as this benchmark model. The origin of this difference with \eqref{DPnper} lies in the subleading corrections, that exactly become increasingly important when $c$ is small (the effective expansion parameter being $1/cN$). In other models, like hybrid inflation \cite{Linde:1993cn}, which end by the action of a transverse symmetry breaking field, the excursion can be even smaller, and still satisfy Planck and BICEP2. We will discuss a similar phenomenon in the next subsection.

\subsection{Sub-Planckian field ranges}

 We now take a  different approach within the logarithmic class of models, in order to illustrate the possibility of obtaining smaller field ranges as compared to the benchmark model of quadratic inflation. The idea is to reproduce the same observational predictions in terms of $(n_s,r)$ by fixing $\tilde N$ (in what follows, we assume $\tilde N=60$) and letting $q$ vary as a function of $p$ through the relation \eqref{deg3}.
  
Once again, the inflationary field range $\Delta \phi$ can be determined by integrating \eqref{dphi1} numerically. We find a striking difference between values of $p$ that are larger or smaller than around $1.1$.

We find that setting the end of inflation by $\epsilon_1=1$ turns out to be possible only for $p$ not exceeding a value around $1.1$. For $p>1.1$ the function $\epsilon_1(N)$, given by \eqref{logep1}, never reaches the unity and, then, a viable inflationary scenario has to be ended through some other mechanism.

\begin{figure}[thb]
\hspace{-3mm}
\begin{center}
\includegraphics[width=8.5cm]{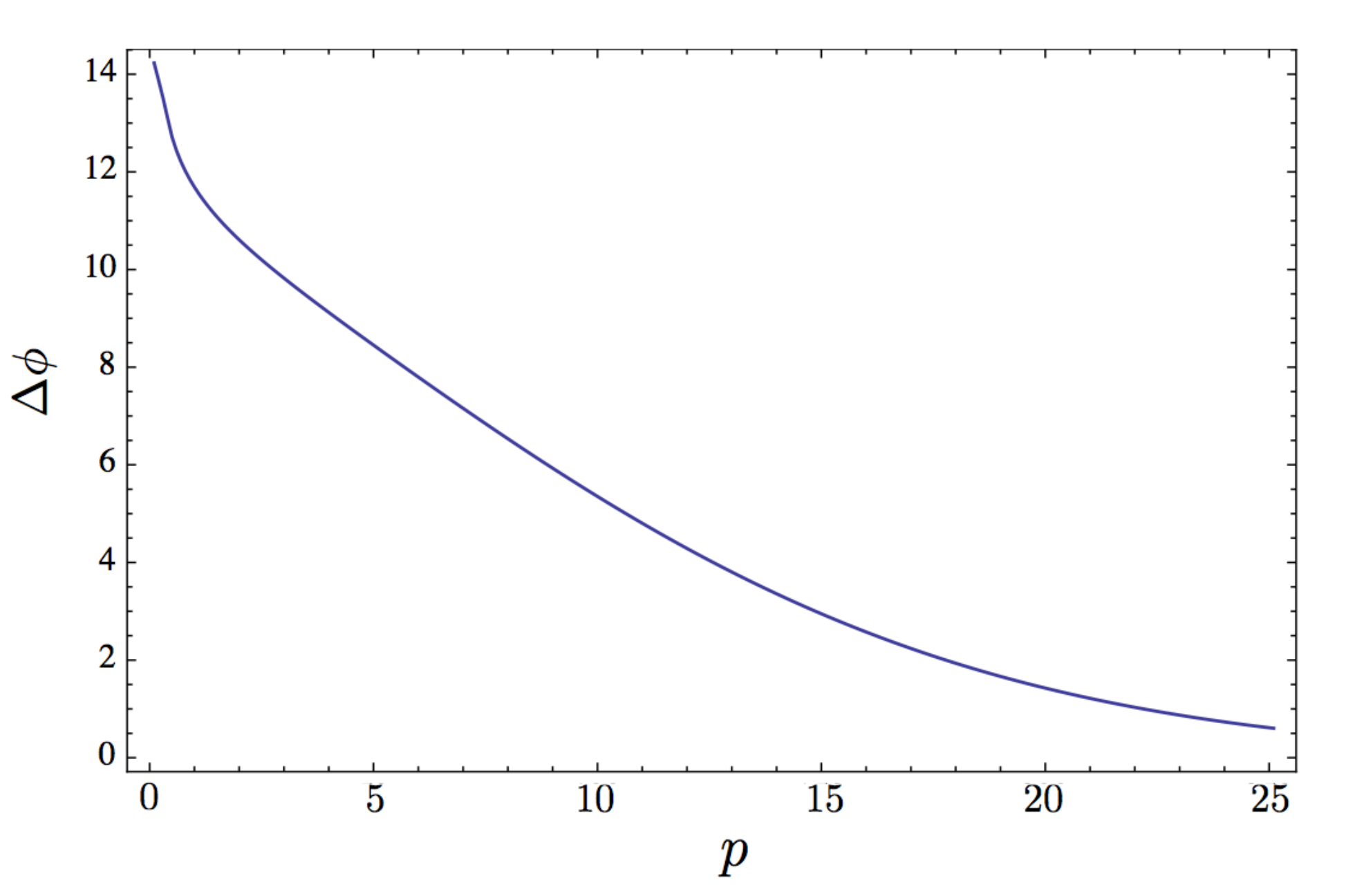}
\caption{Slow-roll field range $\Delta\phi$ as function of $p$. The end of slow-roll inflation is set through the condition $\epsilon_2=1$. Sub-Planckian field ranges can be obtained if inflation ends through a second field or some other mechanism.} \label{FRLogepsJoined}
\end{center}
\vspace{-0.3cm}
\end{figure}

On the other hand, one can still set the end of slow-roll inflation via the condition $\epsilon_2=1$. In this case, the field range is a decreasing function of $p$, as showed in Fig.~\ref{FRLogepsJoined}, and the values of $\Delta\phi$ correspond to the distance which the canonical field $\phi$ travels within the slow-roll approximation. For sufficiently large $p$, such excursion becomes even sub-Planckian. However, note that these models generically would not correspond to slow-roll inflation throughout the whole period of exponential expansion and they would need to end inflation e.g. via a second field, or some other mechanism.

\section{Discussion}
\label{sec:disc}

 In this paper, we have investigated the implications of the CMB data for the inflationary field range. More precisely, we have tried to answer to what extent one can infer $\Delta \phi$ from a measurement of $(n_s, r)$. We have analyzed this question by comparing three different classes of models -- perturbative, logarithmic and non-perturbative -- to the benchmark models of chaotic inflation, with particular attention to the quadratic scenario. 

Surprisingly, we have found that the field range can vary an order of magnitude; while the quadratic model implies $\Delta \phi \approx 14$ in Planck units, the non-perturbative class gives the same observables while $\Delta \phi$ is a factor 11 larger. Moreover, we have identified a continuous degeneracy in the other classes: different one-parameter families of models yield identical $(n_s,r)$ while $\Delta \phi$ spans over a quite large range. Remarkably, $\Delta \phi$ can be increased by exactly the same factor by varying this parameter in both the perturbative and the logarithmic class. Therefore, this constitutes an upper bound for these classes of models.

\begin{figure}[htb]
\hspace{-3mm}
\begin{center}
\includegraphics[width=8cm]{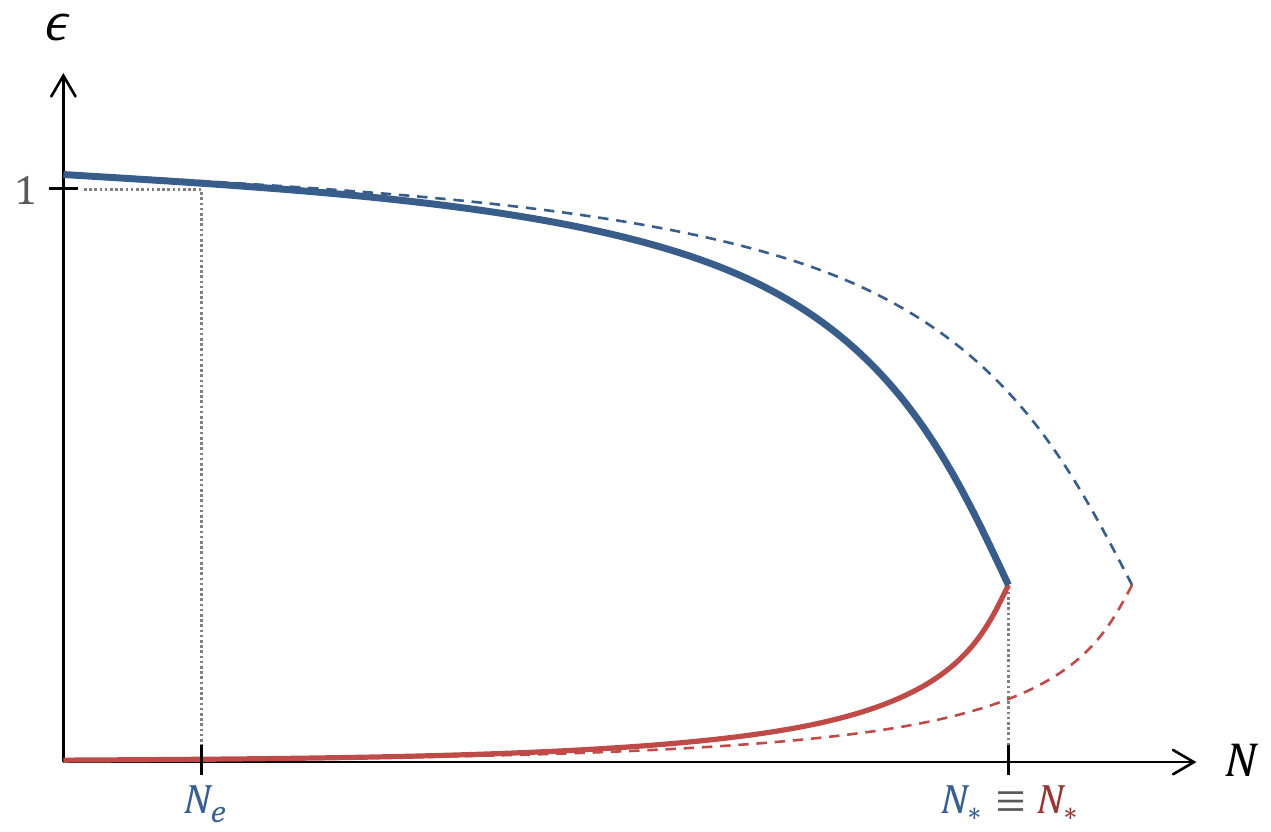}
\caption{Two possible scenarios for the $N$-dependence of $\epsilon$: the solid upper line aims to maximize the field range while the lower minimizes it. The dashed lines correspond to the same scenarios where one increases $N$.} \label{epsilon-N}
\end{center}
\vspace{-0.3cm}
\end{figure}

It might be surprising that there is an upper limit on the field range. After all, we are allowing in principle for an infinite number of e-foldings, hence one would expect it to be possible to hover just below $\epsilon_1 = 1$ for an infinitely long period in terms of $N$; such a scenario is illustrated by the upper line in Fig.~\ref{epsilon-N}. This period would contribute an infinitely large field range $\Delta \phi$ as well. This raises the question: why do we not find such infinitely large field ranges? We suspect that the answer lies in the Hubble flow equations for the slow-roll parameters. For slow-roll inflation, in the approximation where we are only keeping the lowest two slow-roll parameters, these can be written as (where $\epsilon = \epsilon_1$ and $\eta = 2\epsilon_1 + 1/2\ \epsilon_2$)
\begin{align}
\frac{d \epsilon}{dN} = 2\epsilon ( \eta - 2\epsilon) \,, \qquad \frac{d \eta}{d N} = \epsilon (\eta - 3 \epsilon) \,. \label{slowroll}
\end{align}
Note that one cannot have both right-hand sides vanishing at the same time when $\epsilon \neq 0$; therefore it is impossible to keep $\epsilon$ constant over a large range of e-foldings. As a consequence, there is a limit on the number of e-foldings between horizon exit and the end of inflation, for a generic slow-roll model. This is a consequence of the generic lower limit on $d \epsilon / dN$, and translates into a limit on the field range during this period.

Nevertheless, the above discussion constitutes only a generic argument; in fact,  specific and non-generic inflationary models could have yet larger field ranges. Examples are in fact provided by models in the perturbative and the logarithmic classes, with parameter $p<0$. In these models the field range can be arbitrarily large. However, these models are contrary to the large-$N$ approach that we have taken in this paper, where the inflationary period approaches a De Sitter phase as $N$ becomes infinite. For $p$ negative it turns out that one has a cut-off on the number of e-folds preceding the moment of horizon exit. Therefore these do not extend infinitely into the past, approaching a De Sitter phase. In this way it turns out to be possible to evade the generic argument for the upper limit based on \eqref{slowroll} above.

From the perspective of UV-sensitivity, yet more interesting is the question how small $\Delta \phi$ can be, and in particular whether it can reach sub-Planckian values. This point has been discussed in some detail recently in literature. In order to minimise the field range, one would like to have the area under the curve $\epsilon(N)$ in Fig.~\ref{epsilon-N} as small as possible; this case is illustrated by the lower line. Starting at horizon exit, one would therefore need to suppress $\epsilon$ as fast as possible \cite{Hotchkiss:2011gz, Choudhury:2014kma} (see also the very recent discussions \cite{German:2014qza, Gao:2014pca, Bramante:2014rva}). In \cite{Antusch}, however, it was pointed out that this is impossible in the slow-roll approximation, exactly due to Eq.~\eqref{slowroll}; as the right hand sides are bilinear in percent-level slow-roll parameters, these can only vary rather slowly as a function of $N$. This upper bound on the change of $\epsilon$ implies a lower bound on the field range. Amusingly, this is the exact opposite reasoning which led to the large field range discussion above.

The issue of getting a smaller $\Delta \phi$ with respect to the benchmark of the quadratic model
has been investigated explicitly in the different classes. In the single-field slow-roll approximation, we have found that sub-Planckian field ranges do not seem to be possible, in agreement with the recent bound \cite{Antusch}: we could only reduce $\Delta \phi$ by a factor of three, down to $\Delta \phi\approx5$ in Planck units. However, these classes of models allow for a much stronger reduction of the inflationary field range, provided one allows for an alternative end of inflation. In particular, by imposing the condition $\epsilon_2 = 1$, we have found sub-Planckian inflationary trajectories that satisfy all slow-roll single-field requirements. Nevertheless, within these models, the parameter $\epsilon_1$ never reaches the unity and the inflationary expansion needs to be stopped by some other mechanism. Such models could be viable when performing a full fast-roll analysis, or when embedded e.g.~in a multi-field model. Note that this type of multi-field is markedly different from those studied in Ref.~\cite{McDonald:2014oza}; in contrast to that reference, our entire inflationary trajectory is purely single-field, and we only appeal to the second field for a waterfall transition to end inflation.

\section*{Acknowledgments}

We acknowledge stimulating discussions with Daniel Baumann and Gianmassimo Tasinato. We acknowledge financial support from the Madrid Regional Government (CAM) under the program HEPHACOS S2009/ESP-1473-02, from the Spanish MINECO under grant FPA2012-39684-C03-02 and Consolider-Ingenio 2010 PAU (CSD2007-00060), from the Centro de Excelencia Severo Ochoa Programme, under grant SEV-2012-0249, as well as from the European Union Marie Curie Initial Training Network UNILHC PITN-GA-2009-237920.


\bibliography{refs}

\providecommand{\href}[2]{#2}\begingroup\raggedright\begin{thebibliography}{10}

\bibitem{Guth}
A.~H. Guth, ``{The Inflationary Universe: A Possible Solution to the Horizon
  and Flatness Problems}'',
\href{http://dx.doi.org/10.1103/PhysRevD.23.347}{{\em Phys.Rev.} {\bf D23}
  (1981)  347--356}.

\bibitem{Linde81}
A.~D. Linde, ``{A New Inflationary Universe Scenario: A Possible Solution of
  the Horizon, Flatness, Homogeneity, Isotropy and Primordial Monopole
  Problems}'',
\href{http://dx.doi.org/10.1016/0370-2693(82)91219-9}{{\em Phys.Lett.} {\bf
  B108} (1982)  389--393}.

\bibitem{AS82}
A.~Albrecht and P.~J. Steinhardt, ``{Cosmology for Grand Unified Theories with
  Radiatively Induced Symmetry Breaking}'',
\href{http://dx.doi.org/10.1103/PhysRevLett.48.1220}{{\em Phys.Rev.Lett.} {\bf
  48} (1982)  1220--1223}.

\bibitem{Planck}
{\bf Planck} Collaboration, P.~Ade {\em et al.}, ``{Planck 2013 results. XXII.
  Constraints on inflation}'',
\href{http://arxiv.org/abs/1303.5082}{{\tt arXiv:1303.5082 [astro-ph.CO]}}.

\bibitem{bicep}
{\bf BICEP2} Collaboration, P.~Ade {\em et al.}, ``{Detection of B-Mode
  Polarization at Degree Angular Scales by BICEP2}'',
  \href{http://dx.doi.org/10.1103/PhysRevLett.112.241101}{{\em Phys.Rev.Lett.}
  {\bf 112} (2014)  241101},
\href{http://arxiv.org/abs/1403.3985}{{\tt arXiv:1403.3985 [astro-ph.CO]}}.

\bibitem{Audren:2014}
B.~Audren, D.~G. Figueroa, and T.~Tram, ``{A note of clarification: BICEP2 and
  Planck are not in tension}'',
\href{http://arxiv.org/abs/1405.1390}{{\tt arXiv:1405.1390 [astro-ph.CO]}}.

\bibitem{Ashoorioon:2014nta}
A.~Ashoorioon, K.~Dimopoulos, M.~Sheikh-Jabbari, and G.~Shiu,
  ``{Non-Bunch-Davis Initial State Reconciles Chaotic Models with BICEP and
  Planck}'',
\href{http://arxiv.org/abs/1403.6099}{{\tt arXiv:1403.6099 [hep-th]}}.

\bibitem{Ashoorioon:2012kh}
A.~Ashoorioon, P.~B. Dev, and A.~Mazumdar, ``{Implications of purely classical
  gravity for inflationary tensor modes}'',
\href{http://arxiv.org/abs/1211.4678}{{\tt arXiv:1211.4678 [hep-th]}}.

\bibitem{KraussW}
L.~M. Krauss and F.~Wilczek, ``{Using Cosmology to Establish the Quantization
  of Gravity}'', \href{http://dx.doi.org/10.1103/PhysRevD.89.047501}{{\em
  Phys.Rev.} {\bf D89} (2014)  047501},
\href{http://arxiv.org/abs/1309.5343}{{\tt arXiv:1309.5343 [hep-th]}}.

\bibitem{Lyth}
D.~H. Lyth, ``{What would we learn by detecting a gravitational wave signal in
  the cosmic microwave background anisotropy?}'',
  \href{http://dx.doi.org/10.1103/PhysRevLett.78.1861}{{\em Phys.Rev.Lett.}
  {\bf 78} (1997)  1861--1863},
\href{http://arxiv.org/abs/hep-ph/9606387}{{\tt arXiv:hep-ph/9606387
  [hep-ph]}}.

\bibitem{Dai:2014jja}
L.~Dai, M.~Kamionkowski, and J.~Wang, ``{Reheating constraints to inflationary
  models}'', \href{http://dx.doi.org/10.1103/PhysRevLett.113.041302}{{\em
  Phys.Rev.Lett.} {\bf 113} (2014)  041302},
\href{http://arxiv.org/abs/1404.6704}{{\tt arXiv:1404.6704 [astro-ph.CO]}}.

\bibitem{RemCar}
G.~N. Remmen and S.~M. Carroll, ``{How Many $e$-Folds Should We Expect from
  High-Scale Inflation?}'',
\href{http://arxiv.org/abs/1405.5538}{{\tt arXiv:1405.5538 [hep-th]}}.

\bibitem{Mukhanov}
V.~Mukhanov, ``{Quantum Cosmological Perturbations: Predictions and
  Observations}'', \href{http://dx.doi.org/10.1140/epjc/s10052-013-2486-7}{{\em
  Eur.Phys.J.} {\bf C73} (2013)  2486},
\href{http://arxiv.org/abs/1303.3925}{{\tt arXiv:1303.3925 [astro-ph.CO]}}.

\bibitem{Roest}
D.~Roest, ``{Universality classes of inflation}'',
  \href{http://dx.doi.org/10.1088/1475-7516/2014/01/007}{{\em JCAP} {\bf 01}
  (2014)  007},
\href{http://arxiv.org/abs/1309.1285}{{\tt arXiv:1309.1285 [hep-th]}}.

\bibitem{GarRoe}
J.~Garcia-Bellido and D.~Roest, ``{The large-N running of the spectral index of
  inflation}'', \href{http://dx.doi.org/10.1103/PhysRevD.89.103527}{{\em
  Phys.Rev.} {\bf D89} (2014)  103527},
\href{http://arxiv.org/abs/1402.2059}{{\tt arXiv:1402.2059 [astro-ph.CO]}}.

\bibitem{Boyanovsky:2005pw}
D.~Boyanovsky, H.~J. de~Vega, and N.~G. Sanchez, ``{Clarifying Inflation
  Models: Slow-roll as an expansion in $1/N_{efolds}$}'',
  \href{http://dx.doi.org/10.1103/PhysRevD.73.023008}{{\em Phys.Rev.} {\bf D73}
  (2006)  023008},
\href{http://arxiv.org/abs/astro-ph/0507595}{{\tt arXiv:astro-ph/0507595
  [astro-ph]}}.

\bibitem{Antusch}
S.~Antusch and D.~Nolde, ``{BICEP2 implications for single-field slow-roll
  inflation revisited}'',
  \href{http://dx.doi.org/10.1088/1475-7516/2014/05/035}{{\em JCAP} {\bf 1405}
  (2014)  035},
\href{http://arxiv.org/abs/1404.1821}{{\tt arXiv:1404.1821 [hep-ph]}}.

\bibitem{Linde:1983gd}
A.~D. Linde, ``{Chaotic Inflation}'',
\href{http://dx.doi.org/10.1016/0370-2693(83)90837-7}{{\em Phys.Lett.} {\bf
  B129} (1983)  177--181}.

\bibitem{STEG}
D.~J. Schwarz, C.~A. Terrero-Escalante, and A.~A. Garcia, ``{Higher order
  corrections to primordial spectra from cosmological inflation}'',
  \href{http://dx.doi.org/10.1016/S0370-2693(01)01036-X}{{\em Phys.Lett.} {\bf
  B517} (2001)  243--249},
\href{http://arxiv.org/abs/astro-ph/0106020}{{\tt arXiv:astro-ph/0106020
  [astro-ph]}}.

\bibitem{STE}
D.~J. Schwarz and C.~A. Terrero-Escalante, ``{Primordial fluctuations and
  cosmological inflation after WMAP 1.0}'',
  \href{http://dx.doi.org/10.1088/1475-7516/2004/08/003}{{\em JCAP} {\bf 0408}
  (2004)  003},
\href{http://arxiv.org/abs/hep-ph/0403129}{{\tt arXiv:hep-ph/0403129
  [hep-ph]}}.

\bibitem{GBRSZuni}
J.~Garcia-Bellido, D.~Roest, M.~Scalisi, and I.~Zavala, ``{The Lyth Bound of
  Inflation with a Tilt}'',
\href{http://arxiv.org/abs/1408.6839}{{\tt arXiv:1408.6839 [hep-th]}}.

\bibitem{Liddle:2003as}
A.~R. Liddle and S.~M. Leach, ``{How long before the end of inflation were
  observable perturbations produced?}'',
  \href{http://dx.doi.org/10.1103/PhysRevD.68.103503}{{\em Phys.Rev.} {\bf D68}
  (2003)  103503},
\href{http://arxiv.org/abs/astro-ph/0305263}{{\tt arXiv:astro-ph/0305263
  [astro-ph]}}.

\bibitem{Freese:1990rb}
K.~Freese, J.~A. Frieman, and A.~V. Olinto, ``{Natural inflation with pseudo -
  Nambu-Goldstone bosons}'',
\href{http://dx.doi.org/10.1103/PhysRevLett.65.3233}{{\em Phys.Rev.Lett.} {\bf
  65} (1990)  3233--3236}.

\bibitem{Linde:1993cn}
A.~D. Linde, ``{Hybrid inflation}'',
  \href{http://dx.doi.org/10.1103/PhysRevD.49.748}{{\em Phys.Rev.} {\bf D49}
  (1994)  748--754},
\href{http://arxiv.org/abs/astro-ph/9307002}{{\tt arXiv:astro-ph/9307002
  [astro-ph]}}.

\bibitem{Hotchkiss:2011gz}
S.~Hotchkiss, A.~Mazumdar, and S.~Nadathur, ``{Observable gravitational waves
  from inflation with small field excursions}'',
  \href{http://dx.doi.org/10.1088/1475-7516/2012/02/008}{{\em JCAP} {\bf 1202}
  (2012)  008},
\href{http://arxiv.org/abs/1110.5389}{{\tt arXiv:1110.5389 [astro-ph.CO]}}.

\bibitem{Choudhury:2014kma}
S.~Choudhury and A.~Mazumdar, ``{Reconstructing inflationary potential from
  BICEP2 and running of tensor modes}'',
\href{http://arxiv.org/abs/1403.5549}{{\tt arXiv:1403.5549 [hep-th]}}.

\bibitem{German:2014qza}
G.~German, ``{On the Lyth bound and single field slow-roll inflation}'',
\href{http://arxiv.org/abs/1405.3246}{{\tt arXiv:1405.3246 [astro-ph.CO]}}.

\bibitem{Gao:2014pca}
Q.~Gao, Y.~Gong, and T.~Li, ``{The Modified Lyth Bound and Implications of
  BICEP2 Results}'',
\href{http://arxiv.org/abs/1405.6451}{{\tt arXiv:1405.6451 [gr-qc]}}.

\bibitem{Bramante:2014rva}
J.~Bramante, S.~Downes, L.~Lehman, and A.~Martin, ``{Clearing the Brush: The
  Last Stand of Solo Small Field Inflation}'',
  \href{http://dx.doi.org/10.1103/PhysRevD.90.023530}{{\em Phys.Rev.} {\bf D90}
  (2014)  023530},
\href{http://arxiv.org/abs/1405.7563}{{\tt arXiv:1405.7563 [astro-ph.CO]}}.

\bibitem{McDonald:2014oza}
J.~McDonald, ``{Sub-Planckian Two-Field Inflation Consistent with the Lyth
  Bound}'',
\href{http://arxiv.org/abs/1404.4620}{{\tt arXiv:1404.4620 [hep-ph]}}.

\end{thebibliography}\endgroup
\bibliographystyle{utphys}

\end{document}